\relax
\documentstyle[12pt]{article}
\begin{document}
\bibliographystyle{plain}
\title{ AHARONOV-BOHM EFFECT IN LUTTINGER LIQUID AND BEYOND* }
\author{F.V. Kusmartsev}
\maketitle

\begin {verse}

L.D.Landau Institute for Theoretical Physics\\
Moscow, 117940, GSP-1, Vorobiew. Sh. 2,V-334, Russia*

\ \\

and

 \vspace{.3 in}
     {\it Institute for Solid State Physics,University of Tokyo}\\
        {\it Roppongi,Minato-ku,Tokyo 106 } \\

\ \\

e-mail  FEO@jussi.Oulu.Fi
\end{verse}

\ \\
    PACS numbers: $72.10.-d$,  $05.30.Fk$,  $73.20 Dx$,  $75.20 En.$
\ \\

*)will appear in JETP Lett. {\bf 60}, N-9 (1994)

\vfill
\eject

\begin{abstract}
    In  systems with the spin-charge separation, the
period of the Aharonov-Bohm (AB) oscillation becomes
half of the flux quantum. This effect is  at least related to
the fact that for the creation of the holons (spinons) are needed
two electrons. The effect is illustrated on the example of
the Hubbard Hamiltonian with the aid of the bosonization including
topological numbers and exists also in the Luttinger liquid on two chains.
The relation
to a fractional $1/N$ AB effect, which can be associated with a modified
Luttinger liquid, is discussed.
\end{abstract}

\ \\

\ \\
\ \\

In  practically  all  1D strongly-correlated electron systems there exists the
phenomenon known as "spin-charge" separation \cite {And}. It was also recently
argued\cite{Puti} that
the spin charge separation is not only inherent to 1D, but also occurs in
the two-dimensional systems related
to HTSC.  The degrees of freedom associated with the single electron are split
into two independent
 spin and charge degrees of freedom associated with  single particle gapless
excitations: spinons and holons as in $1D$ Luttinger
 liquid\cite{Halda}.

 We show  \cite{KusTreste}that the properties of strongly-correlated systems
are associated with a new type of AB effect,
namely, the period of the AB effect decreases and becomes
half of the period of the AB oscillations for the free electrons\cite{Flux}.
This is valid for all  systems where the spin-charge separation  exists. The
spinon, as well as the holon excitation, is created by two single electron
operators associated with the spin and charge density fluctuations, which is
also the reason why the period of the AB oscillation becomes halves.   With
holons as well  as with  spinons two types of topological numbers are
associated bound with
some selection rules defined by the
parity of the total number of electrons\cite {Halda}.
 As a result all properties are parity dependent and the parity effect exists
in the Luttinger liquid of spinful electrons.
However, the period of the oscillation for the Hubbard ring in the limit of
$U\rightarrow \infty$ decreases  $N_e$ times, where $N_e$ is the number of
electrons on the ring \cite {KusHub}\cite{KWKT}.  The other important feature
of this effect is the absence of the parity effect, which  exists
for free electrons \cite {Flux} \cite {Loss} as well as for interacting
fermions \cite {Flux} \cite {Leget} \cite {Loss2}. The absence of the
parity effect  is also connected with the $1/N_e$  decrease of the
AB period. The system in the
strong-coupling regime can be described by a modified Luttinger liquid.

 To illustrate the decrease of the AB period we consider the Hubbard
Hamiltonian:
\begin{equation}
H = t\sum_{i,\sigma} (a_{i\sigma}^+ a_{(i+1)\sigma} +h.c.) +U
\sum_i n_{i\uparrow} n_{i\downarrow}
\end{equation}
where $t$ and $U$ are hopping integral and the constant of the on-site
Coulomb interaction between electrons, respectively.
   First, we  go to the continuum limit and then apply the bosonization \cite
{Luth}, where the  Hamiltonian is \cite {Luth}
\begin{eqnarray}
H_e = it \sin k_F \sum_{s} \int_0^L dx (\Psi_{s-} (x) \partial_x \Psi_{s-}(x) -
\Psi_{s+}(x) \partial_x \Psi_{s+}(x)) +
\nonumber \\
 U \int^{L}_{0} dx [ :j_{0\uparrow} : : j_{0\downarrow} : +
\Psi_{\uparrow+}^+ (x) \Psi_{\uparrow-}^+(x) \Psi_{\downarrow+}(x)
\Psi_{\downarrow-}(x) + h.c.]
\end{eqnarray}
where $\Psi_{s\pm}$ are left and right movers and
$j_{0\uparrow} = \Psi_{\uparrow+}^+ (x) \Psi_{\uparrow+} (x) +
\Psi^+_{\uparrow-} (x) \Psi_{\uparrow-}(x)$.
The analogous expression is written for  the current
 $j_{0\downarrow}$ of  down-spin fermions.

   We take into account the periodical boundary  (PB),  and twisted boundary
(TB) conditions, when the Hubbard ring is located in a transverse magnetic
field. In both cases the fermion field $\Psi_{\beta\alpha} (x)$ can be
represented as:$\Psi_{\beta\pm} (x) = \frac{1}{\sqrt{2\pi\alpha}} \exp (\pm i
\sqrt{4\pi} \Phi_{\beta\pm} (x)),$
where $\alpha$ is the cut parameter and the boson fields $
\Phi_{\beta\pm} $ for PB conditions can be represented as :
$\Phi_{\beta\pm} (x) = \Phi_\beta(x) \pm \int_{-\infty}^x \pi_\beta (x^\prime)
dx^\prime .$ Here $\pi_\beta (x)$ is the conjugate variables to $\Phi_\beta$.
In terms of these fields the Hubbard Hamiltonian takes the form:
\begin{eqnarray}
H =\int_0^L dx (\frac{ t \sin k_F }{2} [\pi_\uparrow^2 +
(\partial_x\Phi_\uparrow)^2 + (\uparrow\rightarrow\downarrow)] +
\nonumber \ \\
U [\frac{\partial_x\Phi_\uparrow\partial_x\Phi_\downarrow}{\pi}
+ \frac{1}{2\pi^2\alpha^2} \cos [\sqrt{4\pi} (\Phi_\uparrow-\Phi_\downarrow)]]
\end{eqnarray}

On the ring the variables $\pi_\beta$ and $\Phi_\beta$ are multi-valued. It is,
therefore,
convenient to decompose them into the single valued variables
and topological quantum numbers, related to the winding numbers on the ring:
\begin{equation}
\Phi_{\beta\pm} (x) = \Phi_\beta(x) \pm \int_{-\infty}^x \pi_\beta (x^\prime)
dx^\prime + (N_\beta \pm J_\beta) \frac{\sqrt{\pi} x}{2 L}
\end{equation}
where the new variables $\pi_\beta (x)$ and $\Phi_\beta$ are  single valued and
 $N_\beta$, $J_\beta$ are topological numbers associated with the charge and
current on the ring. These numbers are connected by the selection rules. These
rules depend on
the parity of the total number of electrons $N_e$.
Imposing the periodical boundary conditions we get the following selection
rules:$
(-1)^{({N_\beta}\pm {J_\beta})}=(-1)^{(N_e-1)},$
which is a simple generalization of the selection rule
for the Luttinger liquid of spinless fermions\cite{Loss2}.
 Implicitly, these selection
rules dictate that if the number of electrons is odd, then the number
$N_{\beta}$ is even and the number $J_{\beta}$ is odd, or the
number $N_{\beta}$ is odd and the number $J_{\beta}$ is even. On the
other hand  if the number of electrons is even, then the number
$N_{\beta}$ is even and the number $J_{\beta}$ is even, or the
number $N_{\beta}$ is odd and the number $J_{\beta}$ is odd.

 For the  case  of TB conditions one can introduce different
flux values for up and down- spin electrons $f_{\beta}$, where the shift
will have the form:
$\Phi_{\beta\pm}(x) \Rightarrow \Phi_{\beta\pm}(x)\pm \sqrt{\pi} f_{\beta}
x/L$.
We separate the theory into two parts, introducing spin and charge fields
$\varphi_s = (\Phi_\uparrow-\Phi_\downarrow)/2$
 and
$\varphi_c = (\Phi_\uparrow+\Phi_\downarrow)/2$,
the fluxes of the electrical and  magnetic field:
$f_s=(f_{\uparrow}-f_{\downarrow})/2$ and $f_c=(f_{\uparrow}+f_{\downarrow})/2$
and the  topological numbers:
$N_s = N_\uparrow -N_\downarrow ,
J_s = J_\uparrow-J_\downarrow,
 N_c = N_\uparrow+N_\downarrow,
 J_c = J_\uparrow+J_\downarrow.$
 In terms of the topological numbers  and the
single valued variables $\pi_\beta$ and $\Phi_\beta$
 the Hamiltonian can be split into two parts $H=H_c+H_s$, where
\begin{eqnarray}
H_c = A_c\int_0^L dx [\pi_c^2 + (\partial_x\varphi_c)^2 ]+ \frac{A_c\pi} {16 L}
[(J_c + 4f_c)^2 + N_c^2],
\label{charge-Ham} \ \\
H_s = A_s \int_0^L dx [ \pi_s^2 + (\partial_x \varphi_s)^2] +
\frac{A_s\pi}{16L}  [(J_s +4f_s)^2 + N_s^2] +  \nonumber \ \\
\frac{U}{2\pi^2\alpha^2} \int_0^L dx \cos [ \sqrt{16\pi} (\frac{\varphi_s}{A_s}
+ \frac{\sqrt{\pi} N_s x}{4 A_s L})]
\label{spin-Ham}
\end{eqnarray}
associated with the charge and spin degrees of freedom, respectively and
 $A_{c/s}^2 = t \sin k_F\pm U/\pi$. The choice of integer numbers $J_c$, $N_c$,
$J_s$ and $N_s$
is dictated by the selection rules described above. For example, if the
number of electrons $N_e$ is odd, then this means that the numbers
$N_{\uparrow}$ and $N_{\downarrow}$ have different parity, i.e. one
of this numbers is odd the other is even, since
$N_c=N_e=N_{\uparrow}+N_{\downarrow}$. This will mean that the numbers
$J_{\uparrow}$ and $J_{\downarrow}$ have different parity, too and
the number $J_c$ is odd. The fact that the Hamiltonian for charge degrees
of freedom is split into two parts and
the number $J_c$  consists of the sum of the two topological numbers
$J_\uparrow$
and $J_\downarrow$ is the reason why the "holon" Hamiltonian has the flux
period $f_T=\frac{1}{2}$ and not conventional
$f_T=1$.

In the case when $N_e$ is even, the selection rules give that
  the numbers
$N_{\uparrow}$ and $N_{\downarrow}$ have the same parity.
 This means that the  AB  effect is  half-flux
quantum periodic and  the energy-flux dependence is described
by parabolic segments with the minima located at the flux equal
to integers and half-odd integers ( see, eq.(\ref{charge-Ham})).
Thus, there occurs a new parity effect, where there is a difference
in  the behavior  for the odd and the even numbers
of electrons, i.e. there is a shift in the energy-flux dependence
 by a quarter of the elementary flux
quantum. This is in  contrast with the parity effect for spinless fermions
\cite{Flux},\cite{Loss},\cite{Leget}, where the shift is by a half of the flux
quantum. A similar situation occurs for an Aharonov-Cashier effect.

To proceed with the calculation of the current of the Hubbard ring we transfer
our problem into the Lagrangian formalism. We drop the irrelevant spin degrees
of the freedom and
consider only the holon Lagrangian $L_c$ and the action $S_c$.  In the
Lagrangian formalism our fields $\varphi_c$ will depend on space and time
variables $\varphi_c=\varphi_c(x,t)$ and will satisfy PB
 conditions for both variables.
 The multi-valued field $\varphi_c(x,t)$ can be
split into single valued  field $\tilde{\varphi}_c(x,t)$ and terms related
to the winding numbers $n$ and $ m$ with the aid of the relation:
$\varphi_c(x,t)=\tilde{\varphi}_c(x,t)+ \sqrt{\pi} x n/(2L) +\sqrt{\pi} t
m/(2L)$.
Therefore, in the Lagrangian for the charge degrees of the freedom
\begin{equation}
L_c= -\int_0^L dx [ {\dot \varphi}^2_c/ (4A_c) +A_c (\partial_x \varphi_c)^2] +
i\frac{\sqrt{\pi}}{4  L} (J_0+4    f) \int_0^L {\dot\varphi} dx,
\end{equation}
the single valued field
$\tilde{\varphi}_c(x,t)$ can be separated.
The contribution of the orbital motion into the partition function $Z$ of the
ring can be calculated with the aid of the continual integral
over the single valued field $\tilde{\varphi}_c(x,t)$ and sums over the
winding numbers $n$ and $m$
 \cite {Zamol}\cite {Krive}:
\begin{equation}
Z_c = \int D\tilde{\varphi_c} \sum_{n,m} \exp[-S_c(\tilde{\varphi_c},J_0,n,m)]
\label{part-func}
\end{equation}
where the action has the form:
\begin{equation}
S_c = \int_0^L \int_0^\beta dx dt [\dot{\varphi_c^2}/(4 A_c) + A_c
(\partial_x \varphi_c)^2 -i\sqrt{\pi}\frac{\dot{\varphi_c}(J_0+4f)} {4L }]
\end{equation}
 After the summation over the winding
numbers the partition function $Z_c$ takes the form:
$Z_c = Z_0
 \Theta_3 (z_J,q_J)\Theta_3(z_m, q_m)$
where $\Theta_3 (x,y)$ is the theta function and $Z_0$ is the partition
function, associated with the single valued field $\tilde{\varphi_c}$;
$ z_J = \frac{(J_0+4f) \pi }{16 }$,  $q_J = \exp (-\frac{\pi L}{16 A_c
\beta})$, $z_m = 0$, $\quad q_m =\exp (-\frac{\pi A_c \beta} {4 L})$

 One  can derive the low and high temperature asymptotic of the free energy
$F=-T\log Z_c$. In the case of the low temperature limit
$\beta\rightarrow\infty$ we have
\begin{equation}
\Delta F = \frac {\pi A_c (J_0+4f)^2 } {16 L}
\label{free-0}
\end{equation}
 which is
 a flux dependent term  of eq.(\ref{charge-Ham}),
where $J_0$ is even or odd, which corresponds to even or odd number of
particles on the ring respectively. In the case  when  ${\pi L}/{(16 A_c
\beta)}>>1$ or $\beta\rightarrow 0$,
the contribution of the orbital motion is:
\begin{equation}
\Delta F = -2T \exp(-\frac{\pi L T }{16 A_c})\cos(\frac {\pi (J_0+4f) }{8})
\label{high-t-lim}
\end{equation}

Recently, after several beautiful experimental works\cite{22} an
enormous theoretical attention has been devoted to the problem of the
persistent
current( see, Refs.\cite{KWKT}\cite{Weis} and references there).
The problem was stimulated by the discrepancy between the amplitude of
the current estimated theoretically and experimental observations.
The experiments indicate that this amplitude is about several orders
larger than the theories predict.

  The persistent current at zero temperature is equal to
\begin{equation}
J_p = -\partial F/\partial f= -2\pi \frac{V_F}{L} (f + \frac{J_0}{4})
\label{curr-0}
\end{equation}
where $-\frac{1}{4} -\frac{J}{4} \leq f \leq -\frac{J}{4} + \frac{1}{4}$
and $V_F = A_c$ which increases with $U$. This means the
enhancement of the current with electron-electron interaction.
   At high temperatures this enhancement even  larger:
\begin{equation}
J_p = -\pi T \exp({-\frac{\pi LT}{16V_F}}) \sin
[\frac{\pi}{2}(f+\frac{J_0}{4})]
\label{curr-high-t-lim}
\end{equation}
Because of the exponential prefactor the current is strongly reduced
with the temperature but increases exponentially with $U$.
 The characteristic temperature, where the
current is still visible is about
$T_c \sim {V_F}/{L}$, which is nothing but the inter-level distance of the size
quantization.
The described enhancement is not consistent with arguments
of the Ref.\cite{Mull} but agrees with numerical simulations \cite{Avis2}.

 In the strong-coupling limit $U \rightarrow \infty$  the problem  can
be diagonalized with the aid of the Bethe ansatz the spectrum obtained
originally by the author \cite {KusHub} has the form:
\begin{equation}
K_n = \frac{2\pi I_n}{L} + \frac{2\pi}{L} \frac{\sum J_{\alpha}}{N} +
\frac{2\pi f}{L} \label{1}
\end{equation}
 The (half)
integers $I_n$ and $J_{\alpha}$ are holon and spinon quantum numbers,
respectively. If  we
introduce the notations $\phi = \sum {J_{\alpha}}/{N}$
then this equation looks like the
spectrum of spinless fermions in the  flux $f+\phi$.
 In the continuum limit for this spectrum one can write  an effective
Hamiltonian of spinless fermions:
\begin{equation}
H =  \int_0^L [\psi^+ (x) ({K_{f\phi}^2} - k_F^2) \psi(x)] dx
\label {2}
\end{equation}
where $k_F=\pi N/L$, and $K_{f\phi}=K + \frac{2\pi\phi}{L} +
\frac{2\pi f}{L}$ and $K$ is a momentum operator.

 For the comparison of the weak and strong-coupling cases we
represent the holon Hamiltonian (\ref{2}) in the bosonized form.
 With the aid of Loss result\cite{Loss2}, the Hamiltonian of the charge degrees
of freedom takes the form
\begin{equation}
H = V_F \int_0^L [ \pi^2 + (\partial_x \varphi)^2 ] dx +
\frac{V_F\pi}{L}[N^2 + (J+2\phi+ 2f)^2]
\end{equation}
 In  comparison with  eq.(\ref{charge-Ham}), there appears
 the fictitious flux $\phi$, having fractional values $\phi = {p}/{N}$.
 Without external magnetic field the selection rules  have the form:
$(-1)^{N+J} = (-1)^{N_e+1+\phi}$, where the value $\phi$ can be equal to $0$ or
$1$.
The latter means that the topological numbers $N$ and $J$, which are
in the Luttinger liquid coupled, now become decoupled. The latter
came from the fact that the parity of $N_e$ plays no role, since we
can change  the value $\phi$ from 0 to 1 and the value $J$ by 2
without change of the energy. This indicates the violation
of the conventional Luttinger liquid where the topological
numbers $N$ and $J$ are coupled by the parity of  $N_e$.

The parity effect appears at  a  finite  $U$.
Then the solution has a  structure
similar to eq.(\ref{1}) plus the energy of the spin-wave excitations.
Therefore,
 for an odd number of particles in the bosonized form the Hamiltonian is
\begin{equation}
H= V_F  \int_0^L [ \pi^2 + (\partial_x\varphi)^2] dx + \frac{V_F\pi}{L}
[N^2 + (J + 2\phi+2f)^2] +
 \frac{V^2_F}{LU} \mid \sin{2\pi}\phi\mid \label{Ham-finU}.
\end{equation}
One sees that in addition to $N$ and $J$ the topological quantum
numbers, there appears a new term, which is an internal energy of the
field $\phi$ and the energy of the spin-wave excitations.
Now the value of the field $\phi$ can
 not take any rational number, i.e. the finite $U$ lifts the
degeneracy and the parity effect appears.

For an even number of
particles we must change in eq.(\ref{Ham-finU}) the $\sin 2\pi \phi$ to
$\cos 2\pi \phi$. At zero external magnetic flux the
selection rules take the conventional form $(-1)^{N+J} = (-1)^{N_e-1}$, which
dictates that  $\phi=0$. This corresponds to the maximum of the spin-wave
excitation spectrum, i.e. with the external magnetic flux $f$
there appears the spin-wave excitations (nonzero $\phi$, compensating the $f$),
 which  again indicates  a violation of the conventional Luttinger liquid
properties, where the field $\phi$ does not exist.  This question needs special
attention.

Thus, in  weak coupling the magnetization
 is a half flux periodical function. The amplitude of the oscillations
 increases with $U$. Whenever the spin and
charge degrees of freedom are separated and  composite particles (they are the
"holons and spinons" in the considered case ) are created,
 half-flux periodical oscillations of the AB type occur
(see, for comparison, Ref. \cite{Fuji}). Therefore, the period of
 the AB oscillations in any strongly correlated systems
will always decrease.

This effect does not exist  for interacting spinless fermions on a single
channel ring\cite{Flux}\cite{Bouz}. But the effect arises when
the ring consists of two or many chains\cite{KLN}.
The reason for the effect is similar to that
for the Hubbard ring. One  can prove exactly that the AB effect
will have here the period of the half- flux quantum for any interactions
which are no larger than the Fermi energy. When the interaction is
comparable with
the Fermi energy the continuum approach is not applicable and there can occur
the fractional $1/N_e$  AB  effect\cite{KusHub} or a fractional $M/N_e$
 AB  effect\cite{KusT}.
Thus, if in the real HTSC materials in a normal state there occurs spin-charge
separation , the AB effect
must have a half-flux quantum period in the units of the elementary flux
quantum.
To observe such predictions in HTSC
 might be a good challenge for experimentalists.

I thank  Yu Lu , Y. Avishai, A. Luther, A. Nersesyan,
M. Takahashi, I. Krive, T. Ando, S. Katsumoto, M. Kohmoto,
M. Ueda, K.  Kawarabayashi,  S.M. Manning, F. Assaad
and M. Kohno for useful discussions,
 Ministery of Education, Science and Culture of Japan and
International Centre for Theoretical Physics for support and ICTP and ISSP for
the hospitality.


\begin{thebibliography}{99}
\bibitem{And} P.W.Anderson, Phys.Rev.Lett. 67 (1991) 3844
\bibitem{Puti} W.O.Putikka, R.L. Glenister, R.R.P. Singh and H.
Tsunetsugu, Phys. Rev. Lett.{\bf73}, 170   (1994);
 Y.C. Chen, A. Moreo, F. Ortolani, E.Dagotto
and T.K. Lee, Phys. Rev. B50, 655 (1994)
\bibitem {Halda} F.D.M. Haldane,  J. Phys.C {\bf14}, 2585 (1981)
\bibitem{KusTreste} The preliminary results of this work
has been presented in F.V.Kusmartsev," AB effect in Luttinger liquid", Preprint
ICTP, Trieste, ICTP (1992)
\bibitem{Flux} F.V.Kusmartsev JETP Lett. 53(1991)28 ; Phys. Lett.A 161 (1992)
433
\bibitem{KusHub} F.V.Kusmartsev, J. Phys. Cond. Matt. 3 (1991) 3199

\bibitem{KWKT} F.V.Kusmartsev, J. Weisz, R. Kishore and Minoru Takahashi,
in Technical report of ISSP, Ser.A, N2772, January 1994
and Phys. Rev. {\bf B49}, 16 234, 1994.

\bibitem {Loss} D.Loss and P. Goldbart, Phys.Rev. B {\bf43}, 13762 (1991)
\bibitem {Leget} A.J. Legett, in NATO ASI, B251, Plenum, NY, 1991, 297
\bibitem {Loss2} D.Loss, Phys.Rev.Lett. {\bf69}, 343 (1992).
\bibitem{Luth} A.Luther and I.Peschel, Phys. Rev. B 9, 2911 (1974)
\bibitem {Zamol} Al. B Zamolodchikov, Nuel. Phys. B, {\bf285}, 481 (1987).
\bibitem {Krive} I.V.Krive and A.S.Rozhavsky, Int. Jour. Mod.Phys. B6
(1992)1255
\bibitem{22} L. Levy et al Phys. Rev. Lett. 64, 2074 (1990),
   V. Chandrasekhar et al ibid 67, 3578 (1991), D. Mailly
et al ibid 70, 2020 (1993)
\bibitem{Weis}
  J. Weisz, R. Kishore and F.V.Kusmartsev, Phys. Rev. {\bf B49}, 8126 (1994)

\bibitem{Mull} A. Mueller-Groeling and H. Weidenmueller, Phys. Rev.
{\bf B49}, 4752 (1994)

\bibitem{Avis2}  Y. Avishai and R. Berkovits, Phys. Rev. Lett. (1994) submitted
\bibitem{Fuji}S. Fujimoto and Norio Kawakami, Phys. Rev. {\bf B46},17 406
(1993)

\bibitem{Bouz}
 G. Bouzerar, D. Poilblanc and G. Montambaux, Phys. Rev.{\bf49}, 8258  (1994)




\bibitem {KLN} F.V.Kusmartsev, A.Luther and A. Nersesyan,
JETP Lett.55
(1992) 724; A.A. Nersesyan, A. Luther and  F.V.Kusmartsev,
 Physics Letters A176 (1993) 363


\bibitem{KusT} F.V.Kusmartsev and Minoru Takahashi,
in Technical report of ISSP, Ser.A, N2992, Juni, 1994
and Phys. Rev.B submitted



\end{thebibliography}
\end{document}